\documentstyle[pre,aps,epsfig,multicol]{revtex}

\begin{document}

\draft

\title{Breakdown of a conservation law in incommensurate systems}
\author{L. Consoli, H. J. F. Knops, and A. Fasolino}
\address{Institute for Theoretical Physics, University of Nijmegen, 
    Toernooiveld 1, 6525 ED Nijmegen, The Netherlands}\date{\today}
\maketitle
\begin{abstract}
We show that invariance properties of the Lagrangian of an incommensurate system,
as described by the Frenkel Kontorova model, 
imply the existence of a generalized angular momentum which is an integral of motion
if the system remains floating. 
The behavior of this quantity can therefore monitor the character of the system 
as floating (when it is
conserved) or locked (when it is not).
We find that, during the dynamics, 
the non-linear couplings of our model cause parametric 
phonon excitations which lead to the appearance of Umklapp terms and to a
sudden deviation of the generalized momentum from a constant value, signalling 
a dynamical transition from a floating to a pinned state. We point out that this 
transition is related but does not coincide with the onset of sliding friction 
which can take place when the system is still floating.  
\end{abstract}

\begin{multicols}{2}

\section{Introduction}
\label{intro}
Measuring friction at an atomic level has recently become experimentally 
possible\cite{exp}. Many studies of the dynamics of appropriate non-linear 
systems aiming at establishing the mechanisms giving rise to energy dissipation 
during the sliding of a body onto a crystalline surface have appeared in the 
literature\cite{Shinjo,Persson,Strunz,Con1}. The Frenkel-Kontorova (FK) model, which describes a 
harmonic chain interacting with a rigid periodic substrate, is particularly 
suitable to study the important case of an incommensurate (IC) lattice
parameter ratio of the contacting surfaces. The present study focuses on the 
effects of discommensuration on the dynamics. It should be kept in mind that
a more realistic study of friction would require an extension to two dimensions.
Coupling to the third dimension can be provided either by an ad hoc damping term or 
by coupling to an elastic medium\cite{Rod}. The ground state properties of this model 
have been thoroughly studied\cite{Aubry}.
At a critical value $\lambda_c$ of the coupling to the external 
potential, the ground state of the system displays a structural transition
(Aubry transition) from a floating to a pinned configuration. Below this 
threshold, the center of mass of the static system can be displaced
on the substrate without energy costs. Therefore, one might expect a frictionless 
regime also in a dynamic situation and a {\it superlubric} regime, where 
the chain would slide indefinitely, has been predicted for this 
case\cite{Shinjo}. In a previous paper\cite{Con1} we have pointed out
that the inherent non-linear coupling of the center of mass (CM) motion to the 
phonons leads instead to an irreversible decay of the CM velocity. 
The essential mechanism for
the transfer of kinetic energy from the center of mass to the internal
vibrations is the parametric resonant excitation of phonons mediated by 
ordinary resonances with phonons related to the modulating potential.

Here we show that this type of mechanism has another important consequence, 
namely it causes the appearance of Umklapp terms, signalling a dynamical 
transition in the system from a floating to a pinned state. We have studied this 
phenomenon by identifying a new quantity, which we call 
generalized angular momentum (GAM), which is an integral of motion only if 
the system is in a floating IC phase, reflecting the invariance of the Lagrangian of the model 
for a phase shift in this state.
We show that this invariance is equivalent to the absence of Umklapp terms. 
By means of numerical simulations we show that the temporal behavior of the GAM 
is a powerful probe both of the (in)commensurability of the groundstate configuration and of 
the dynamical phase in which the system is during motion. 
Simulations where the incommensurate ground state is given an initial velocity
show that the GAM remains conserved up to a well defined time  where a sudden jump
takes place. We have been able to relate this change of behavior from 
conserved to non conserved to the appearance of Umklapp terms. An important finding is 
that this floating-pinned transition does not coincide with 
the onset of friction. It was recently suggested by Popov\cite{Popov} that the appearance
of Umklapp terms, i.e. the conservation of quasi-momentum instead of momentum for
crystalline systems, is the mechanism via which friction occurs in incommensurate
contacts. The present result shows that this is not the only mechanism. 
By monitoring the system via the GAM we can show that
decay of the CM velocity may occur already in the floating phase. 
The onset of friction and the appearance of Umklapp terms are both caused 
by non-linear couplings and resonant phonon excitations in the system but remain 
two distinct phenomena occurring at different times.
  
In Sec. \ref{res} we describe the construction of the GAM 
by deriving it from the Lagrangian for the system
in Fourier space and define conditions under which it is conserved.
In Sec. \ref{num} we present results of
numerical simulations which confirm the validity of our analytical derivation
and underline the usefulness of the GAM to discriminate between commensurate 
and floating IC and pinned IC phases, respectively. 
Subsequently, we examine in Sec. \ref{umkl} the relationship between 
pinning and Umklapp and show the presence of a well-defined transition 
time. In Sec. \ref{concl} we present  conclusions 
and perspectives of this work.    
In the Appendix we provide the reader with an explicit proof that the GAM 
is an integral of motion in absence of Umklapp.

\section{Construction of a Generalized Angular Momentum}
\label{res}

In this section we will construct a generalized angular momentum for the
dynamical FK model, as described in Ref. \onlinecite{Con1}. We remind the reader 
that this model represents a chain of $N$ particles which interact with each other 
via a first-neighbor harmonic potential and are subjected to an external, spatially 
periodic, potential of strength $\lambda$.
The FK Hamiltonian reads:
\begin{equation}
{\cal H}=\sum_{n=1}^N\left[\frac{p_n^2}{2} + \frac{1}{2}\left(u_{n+1}-u_n-l
\right)^2 
+\frac{\lambda}{2\pi}\sin{ (\frac{2\pi u_n}{m} )}\right]
\label{Ham}
\end{equation}
where the $u_n$ are the particle positions and $p_n$ their momenta.
The ratio between the modulation 
period of the external potential $m$ and $l$ (the equilibrium distance 
between the atoms of the chain for $\lambda=0$) is taken to be irrational, i.e. 
the system is incommensurate. In our calculations 
we take $m=1$ and $l=\tau=\frac{\sqrt{5}+1}{2}$ (golden mean). 
In the numerical implementation for a finite system of $N$ particles 
we impose periodic boundary conditions 
\begin{equation}
u_{N+1}=Nl+u_1
\label{pbc}
\end{equation}

This implies that we have to choose commensurate approximants for the 
equilibrium distance $l$. By expressing $l$ as ratio of consecutive Fibonacci 
numbers we obtain approximants which satisfy  
the condition $l \times N = M \times 1$ with $M$ and $N$ integers. 
Let us introduce the modulation wavevector $q=2\pi l=2\pi(M/N)$ and the position and 
momentum of the CM of the chain of atoms:
\begin{equation}
 Q = \frac{1}{N}\sum_n u_n \mbox{,\hspace{1cm}}P = \frac{1}{N}\sum_n p_n 
\label{cmcoord}
\end{equation}
The equations of motion for the deviations $x_n=u_n - nl -Q$ from the 
equilibrium positions in 
the uncoupled chain are then given by: 
\begin{equation}
\ddot{x}_n=x_{n+1}+x_{n-1}-2x_n+\lambda\cos{(qn+2\pi x_n+2\pi Q)}
\label{eom1}
\end{equation}
As noted in Ref. \onlinecite{Con1}, in the weak coupling regime it is useful 
to move to Fourier 
coordinates $x_k=\frac{1}{N}\sum_n e^{-ikn}x_n$ with $k=2\pi K/N$. The phonon dispersion of the
chain for $\lambda=0$ is denoted by $\omega_k\equiv\omega(k)=2|\sin{(k/2)}|$.
The Lagrangian associated with Eq. (\ref{Ham}) in transformed space becomes:
\[
{\cal L} = N\left(\sum_k \left(\frac{1}{2}\dot{x}_k\dot{x}_{-k} - \frac{1}{2}\omega_k^2 x_k x_{-k}\right) + \right.
\]
\[
\left.
\frac{\lambda}{2\pi}\frac{1}{2i}\sum_{m=1}^{\infty}\frac{(i2\pi)^m}{(m!)}\sum_{k_1 
\ldots k_m}\left(e^{i 2\pi Q}x_{k_1}\cdots x_{k_m}\delta_{k_1+\cdots+k_m,-q}\right.\right.
\]
\begin{equation}
\left.\left. - (-1)^m e^{-i 2\pi Q}x_{k_1}\cdots x_{k_m}\delta_{k_1+\cdots+k_m,q}\right)+\frac{1}{2}
(\dot{Q})^2\right)
\label{lagrk}
\end{equation}
It is important to notice that since wave vectors are defined modulo $2\pi$, 
the Kronecker deltas
in Eq. (\ref{lagrk}) should be read as:
\begin{equation}
k_1 + k_2 + \cdots + k_m = q + s\times 2\pi
\label{kron}
\end{equation}
Umklapp is present whenever this relation is satisfied with $s\neq 0$. 
It is clear that the occurrence
of Umklapp depends on the modes $x_k$ which are not negligible, and on the choice of 
the (extended or reduced) Brillouin zone in which $k$ is represented.
It is known that in the groundstate, for a coupling $\lambda$ well below 
the critical value $\lambda_c$, which for this model assumes the value $\lambda_c=0.154...$, the modes 
with wavevector $nq$ have an amplitude which scales as $\lambda^{|n|}$. This number $|n|$ is
therefore a natural label to represent the modes; we define $n(k,q)$ as the 
smallest (in absolute value) number which satisfies:
\begin{equation}
k=n(k,q)q\,\,\mbox{mod}\,(2\pi)
\label{new_k}
\end{equation}
For a finite system with $N$ particles, where $k$ can be represented in 
the reduced Brillouin zone
as $k=K(2\pi/N)$, $K \in (-1/2N, 1/2N]$, this can be rewritten as:
\begin{equation}
K = nM\,\,\mbox{mod}\,(N)\,\,\,\,\mbox{,}\,n\in (-1/2N, 1/2N]
\label{k_K}
\end{equation}

\begin{figure}
\epsfig{file=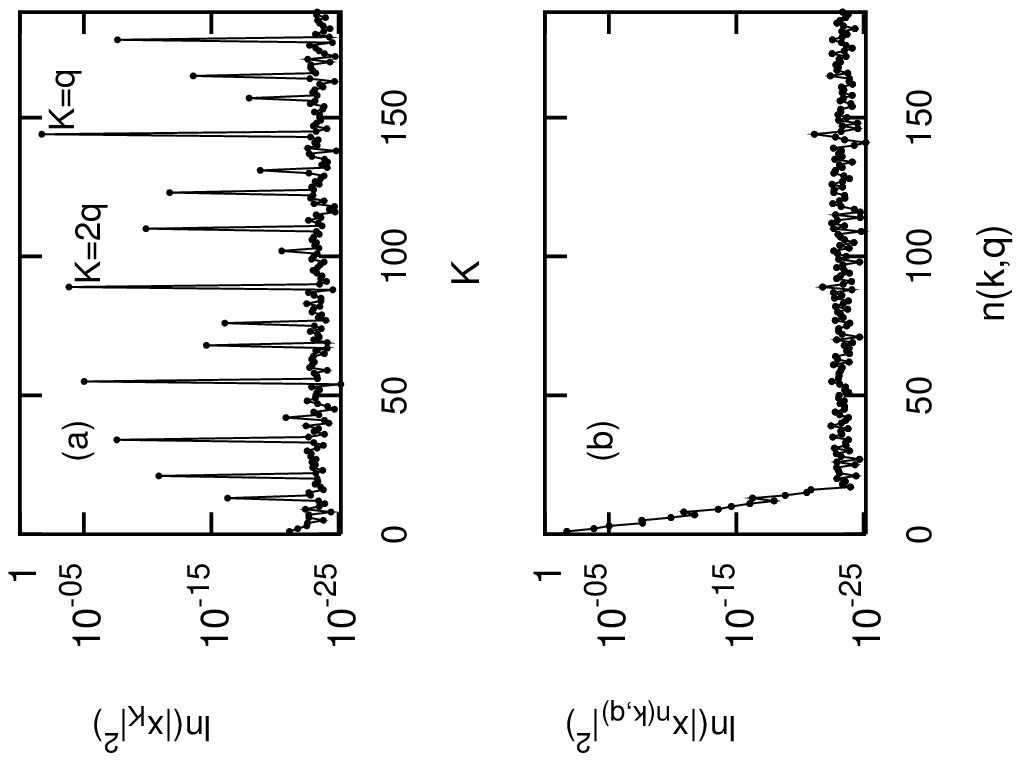, angle=-90}
\caption{FK model for $N=377$, $\lambda=0.05$. (a) Phonon
amplitudes squared plotted as a function of the wavevector $K$, as in Eq. (\ref{k_K}). The first two $nq$ modes are 
explicitly indicated.
(b) Same as in panel (a), relabeled according to Eq. (\ref{new_k}).
Due to finite numerical precision, the exponential decay with $\lambda^{|n|}$ is 
apparent only for the first fifteen modes.}
\label{fig1}
\end{figure}

In Fig. \ref{fig1} we compare the phonon amplitudes for the ground state of the FK model for
$N=377$, $\lambda = 0.05$, plotted 
as a function of the usual wavevector label $K$ (panel (a)), as 
well as reordered according to the label $n$ (panel (b)). 
Note that, due to finite numerical 
precision, the scaling behavior is hidden in numerical noise after the first 
fifteen modes.

The use of $n$ as a mode label makes apparent the fact that there is no Umklapp in 
the ground state of the FK model in the modulated phase 
for $\lambda < \lambda_c$.
In fact Umklapp would imply the presence of a non vanishing term:
\begin{equation}
x_{n_1 q}x_{n_2 q}\cdots x_{n_m q}\,\mbox{;}\,\,\,\,\,n_1 + n_2 + \cdots n_m = sN\,\,\mbox{ with }s\neq 0
\label{umk_term}
\end{equation}
The joint amplitude of this term would be:
\begin{equation}
\lambda^{|n_1| + \cdots |n_m|} \le \lambda^{|s|N}
\label{amp_umk}
\end{equation}
which vanishes for $N\rightarrow\infty$.

The absence of Umklapp terms is directly related to the existence of a free floating phase, which
is a well known invariance property of the FK model. In the present notation it amounts to 
the invariance of the Lagrangian for the transformation:
\begin{eqnarray}
Q & \rightarrow & Q + q\phi / 2\pi \\
x_k & \rightarrow & x_k e^{ik\phi}
\end{eqnarray}
This invariance is related to the existence of a zero-frequency Goldstone mode in the system.
This mode is also often called phason, and should not be confused with the usual acoustic mode
of periodic crystals.

Having found an invariance for the Lagrangian, we can look for the 
conjugate conserved momentum. 
We get:
\begin{equation}
p_{\phi} = \frac{\partial{\cal L}}{\partial\dot{\phi}}= -i\sum_n nqx_{-nq}\dot{x}_{nq}
+\frac{q}{2\pi}\dot{Q} \equiv L + \frac{q}{2\pi}\dot{Q}
\label{gam}
\end{equation}
The quantity $p_{\phi}$ represents a generalized angular momentum (GAM). 
It is important to realize that the invariance of the Lagrangian only holds 
in a subspace of
the full phase space where Umklapp terms can be neglected as it 
is the case for the floating (incommensurate) 
ground state. In order to stress this point, a direct calculation of $\dot{p}_{\phi}$ 
is given in the Appendix, 
showing that the GAM $p_{\phi}$ is an integral of 
motion only if Umklapp is not present.

This quantity is therefore an useful tool to discriminate between
commensurate and incommensurate structures, and floating and locked states.
In the next section we present numerical simulations that we carried 
out for various values of the parameters of the model, showing how
$p_{\phi}$ is a good indicator of the phase in which the system under examination is.

\section{Numerical results}

\subsection{Commensurate vs. incommensurate, locked vs. floating}
\label{num}

\begin{figure}
\epsfig{file=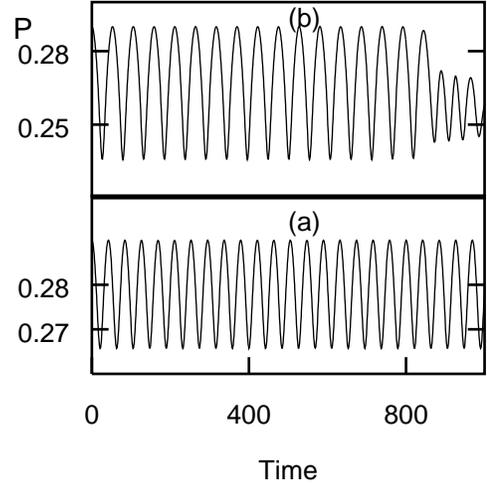, angle=-90}
\caption{Behavior of the CM momentum P for: (a) an incommensurate configuration with $N=144$, $\tau=233/144$, 
$\lambda=0.015$, $P_0=0.29$; (b) a commensurate case with $N=144$, $\tau=5/3$, $\lambda=0.015$, $P_0=0.29$. Note 
the qualitative similarity in the behavior of $P$.}
\label{fig2}
\end{figure}

\begin{figure}
\epsfig{file=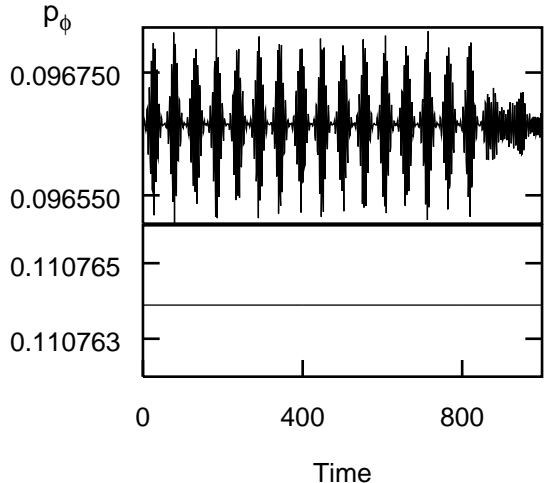, angle=-90}
\caption{Behavior of $p_{\phi}$ for the parameters of the model as described in Fig. \ref{fig2}. (a) Incommensurate 
case: the GAM is constant within numerical precision. (b) Commensurate case: the GAM is 
not conserved. Note the change of scale going from panel (a) to panel (b).}
\label{fig3}
\end{figure}

We have performed numerical simulation in order to study the behavior of the GAM, as 
defined by Eq. (\ref{gam}), integrating by a Runge-Kutta algorithm the $N$ Eqs. (\ref{eom1}).
We assign to the particles of the chain as initial conditions momenta $p_n=P_0$ and positions $x_n(t=0)$ 
corresponding to the ground state. Fig. \ref{fig2} shows simulation results for the same
number of particles $N$ and potential strength $\lambda$, but for a low ($\tau = 5/3$) and a 
high ($\tau = 233/144$) approximant 
to the golden mean $\tau$, producing a commensurate structure 
and an approximate incommensurate one. 
The qualitative behavior of the momentum of the center of mass $P$ is
similar, whereas the behavior of $p_{\phi}$ in Fig. \ref{fig3} is remarkably 
different being conserved only for the case which approximates an incommensurate system.  
This confirms that $p_{\phi}$ can be used as a tool to discriminate 
unambiguously between 
commensurate and incommensurate structures.

\begin{figure}
\hspace*{-1.5cm}\epsfig{file=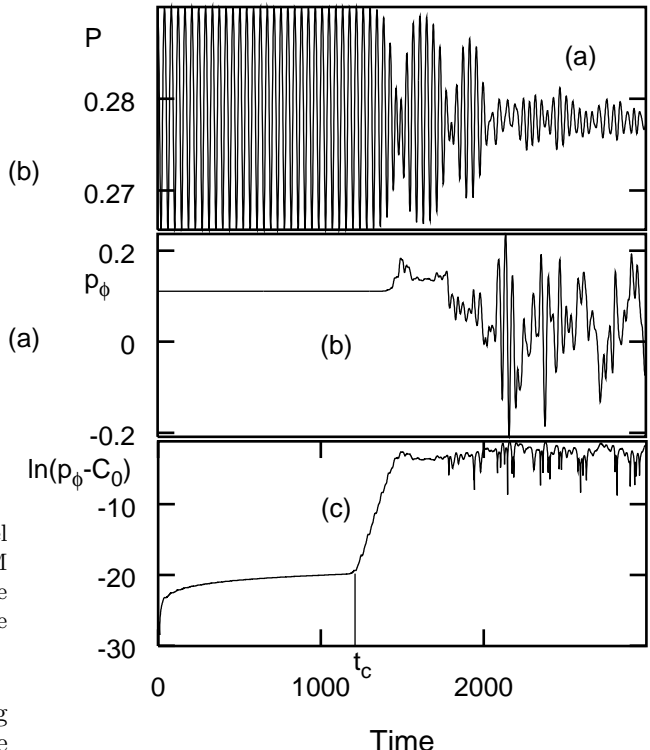, angle=-90}
\caption{$N=377$, $\tau=610/377$, $\lambda=0.015$, $P_0=0.29$. (a) Behavior of the CM momentum P. (b) 
Behavior of $p_{\phi}$. It is possible to see how the GAM stops being conserved. (c) In order
to check if there is a critical time $t_c$, we plot the quantity 
$log(p_{\phi}-C_0)$, where $C_0=p_{\phi}(t=0)$. We show that $t_c$ can be unambiguously identified.
Note that, even if, in this weak coupling case, the decay of the CM coincides with the breakdown of 
conservation of $p_{\phi}$, this is not always the case. See also text and Fig. \ref{fig6}.}
\label{fig4}
\end{figure}

Furthermore, our numerical simulations show a remarkable fact. If we start the simulation with
an incommensurate initial condition, $p_{\phi}$ is indeed conserved, but only up to a critical
time $t_c$, after which it rapidly deviates from its initial conserved value.
This is shown in Fig. \ref{fig4}, where 
we can examine the behavior of $p_{\phi}$ and $P$ in a weak coupling, highly incommensurate ($\tau=610/377$, 
$\lambda=0.015$) case. In order to check that the observed variation of $p_{\phi}$ only sets in 
after a critical time $t_c$, we have analyzed the behavior of the quantity
$log(p_{\phi}-C_0)$, $C_0$ being the value of $p_{\phi}$ at $t=0$. It is evident 
from panel (c) of Fig. \ref{fig4}
that we can identify such a critical time $t_c$ where the GAM has a jump in value of various order of 
magnitude.  Besides, this figure shows that, for $t<t_c$, $p_{\phi}$  is conserved 
within our numerical accuracy, never exceeding variation larger than $10^{-20}$.  
The critical time $t_c$ obviously depends on the coupling strength $\lambda < \lambda_c$ and on the 
initial velocity $P_0$. We are currently investigating the dependence $t_c(\lambda, P_0)$, which turns 
out to be rather intricate, and plan to report on it in the near future. Here, we want to concentrate
on the mechanism that causes the breakdown of the conservation law, leading to a well-defined $t_c$.
We show in the next section that the answer lies 
in the appearance of Umklapp terms, which render the system pinned, thus leading to a 
non conservation of $p_{\phi}$. 

\subsection{The role of Umklapp processes}
\label{umkl}

As we have pointed out in Sec. \ref{res}, $p_{\phi}$ is conserved only in the absence of Umklapp, that 
is to say when terms of the form given in Eq. (\ref{umk_term}) have a vanishing amplitude for 
$N\rightarrow\infty$. As we have seen, this is the case for the groundstate of the incommensurate
system. However, starting from the ground state, this may change during the dynamics due to
parametric resonances. The movement of the CM with velocity $P$ induces a modulation with
frequency $\Omega = 2\pi P$ in the equations of motion of the system. Linear stability 
analysis\cite{Strunz,Con1} shows that a mode $k$ grows exponentially (with rise time $\tau$)
whenever its frequency satisfies:
\begin{equation}
\Omega\simeq\frac{\omega(k) + \omega(mq-k)}{m}
\label{res_rule}
\end{equation}
for some $m$. The $\simeq$ symbol indicates an instability window of relative width $w_m$ 
that scales with $\lambda^{|m|}$.
Suppose now that a mode $x_k$ is unstable. Let $n(k,q)$ be its label according 
to Eq. (\ref{new_k}).
When $n(k,q)={\cal O}(N)$, this mode could lead to Umklapp as soon as its 
amplitude becomes finite.
However, since its initial value is of order $\lambda^{|n|}$, it requires an 
infinite amount of time
(as $N\rightarrow\infty$) to render $\lambda^{|n|} e^{t/\tau}$ finite.
So we have to look for modes $k$ in the instability windows with $n(k,q)$ finite.
Since the mapping $n(k,q)\rightarrow k$ leads to a uniform distribution, one indeed expects 
to find such a $k$ with $|n(k,q)| < 1/w$, where $w=\sum_m w_m$ is the relative width of 
the joint instabilities windows.
Once the amplitude of the mode $k$ starts growing (with a 
behavior given by an exponential law of the 
form: $Ae^{t/\tau}$, with $A$ bounded from below via the upper bound on $n$), also 
modes $k'$ with $n'(k',q) = pn(k,q)\pm 1, p=2,3,...$ start to 
develop via non linear terms
in the equations of motion (see Ref.\onlinecite{Con1} for details), with the form:
\begin{equation}
x_k^{'} \simeq \lambda(Ae^{t/\tau})^p
\label{cascade}
\end{equation}
Umklapp can result when repeating this process ${\cal O}(N)$ times (i.e., when $pn \sim N$) 
still gives a finite result.
From Eq. (\ref{cascade}) it is clear that this will happen when $x_k(t)$ exceeds some threshold 
value,  which is the case for $t$ larger than the critical time $t_c$ 
introduced in Sec. \ref{num}.
\begin{figure}
\hspace*{-1.5cm}\epsfig{file=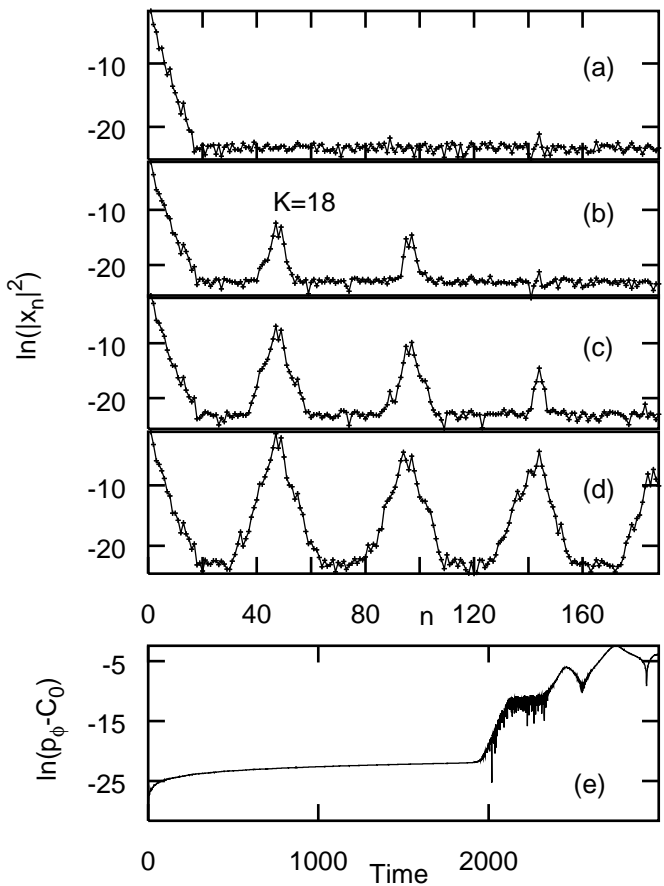, angle=-90}
\caption{$N=377$, $\tau=610/377$,$\lambda=0.05$, $P_0=0.15$. The appearance of Umklapp terms is shown. 
(a) Phonon amplitudes squared at $t=0$. (b) Same at $t=1000$. The mode $K=18\,\,(n=47)$ is starting to 
grow; (c) Same  at $t=1500$. The modes with $n \times 47$, with $n=1,2,3$ are all present, but there 
is still no Umklapp. (d) The mode with $K=18$ has become order unity and Umklapp appears, signalled by
the excited modes at the zone boundary. (e) Time evolution
of $log(p_{\phi}-C_0)$. The appearance of Umklapp in panel (e) corresponds to the breakdown of 
the conservation of the GAM. See also text for a detailed explanation of the Umklapp mechanism.}
\label{fig5}
\end{figure}
Fig. \ref{fig5} shows this mechanism at work: here we have taken $P_0=0.15$, so that 
$\Omega=2\pi P=\omega_q/2$, and $N=377$ $(\tau=610/377)$. For this value of $\Omega$ 
Eq. (\ref{res_rule}) has 
approximate solutions for $n=2$ for $k=K(2\pi/N)$, with $K=18,19$. The corresponding values for
$n(k,q)$ are in the case $n(18,q)=47$ and $n(19,q)=97$. Panel (a) shows that the phonon amplitudes 
at $t=0$ decay indeed exponentially with $n$. Panel (b) shows that the unstable mode $K=18$ with
the lowest value of $n$ starts to grow at $t=1000$. Modes at $n=2 \times 47$ are also present, due to
non linear terms, but quadratically smaller, as explained above. At $t=1500$ modes at $n=3\times 47$ 
become visible, as shown in panel (c). There is still no Umklapp, as can be seen from the absence of
an amplitude at the zone boundary. 
Such a term, corresponding to $n=4\times 47$, finally appears in panel (d), at $t=2000$. 
Indeed, this is
also precisely the time at which $|x_{18}|^2$ becomes order unity and $p_{\phi}$ stops being
conserved (see panel (e)).

The mechanism described above shows that the appearance of Umklapp terms causes a sudden 
transition from a floating to a pinned structure. 
In this respect, this transition represents a dynamical analogue of the Aubry
transition taking place in the static model at $\lambda_c$.
The important difference is that, for the dynamical case, this transition occurs as a 
function of time at all values $\lambda < \lambda_c$. 
 \begin{figure}
\epsfig{file=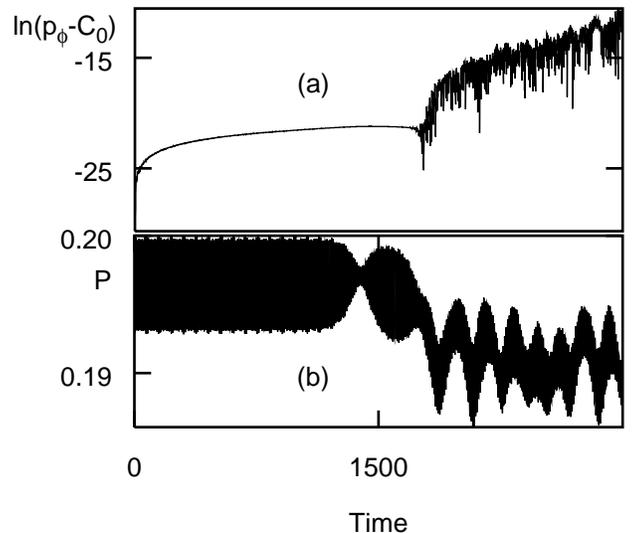, angle=-90}
\caption{$N=144$, $\tau=233/144$, $\lambda=0.05$, $P_0=0.20$. (a) Time evolution of $log(p_{\phi}-C_0)$.
(b) CM momentum $P$. Note that the CM momentum has begun its decay before the deviation of $p_{\phi}$ from
its constant value occurs.} 
\label{fig6}
\end{figure}

Before concluding this section it is important to discuss the relation 
between this "dynamical Aubry" transition described above 
and the onset of friction. The onset of friction is driven by the coupling of the 
CM to  the mode with the modulation wavevector $q$ or its harmonics and consists 
in a special kind of parametric resonances involving more than one phonon and 
where the time dependent driving terms are themselves in resonance\cite{Con1}. 
The appearance of Umklapp requires instead a
phonon with finite amplitude and very special (zone boundary) wavevector.  
We could say that the last process is more difficult to achieve. 
In Fig. \ref{fig6}, it can be seen
that the GAM stays conserved, even after the CM momentum has begun its decay. This means
that there is an interval of time in which the first mechanism is active whereas the 
second has not yet taken place. We can identify
two times: one which characterizes the onset of friction and one which describes the 
pinning of the system. In this respect the dynamical model is much richer than 
the statical one. At the Aubry transition,  the appearance of a static friction 
occurs by definition at the same value of $\lambda$ at which the system gets pinned.

\section{Conclusions}
\label{concl}

We have shown, in the framework of the undamped 1D dynamical FK model, 
that it is possible to
obtain analytical results concerning the existence of a new integral of motion 
which represents a 
generalized angular momentum related to a phase invariance in incommensurate systems, and we 
have confirmed this finding by means of numerical simulations. We have also shown that, during the
dynamics, a breakdown of the conservation of the GAM occurs at a well-defined time, signalling
a dynamical transition from a floating phase to a locked one. We have been able to prove that 
this transition is related to the apperance of Umklapp processes, caused by non linear 
couplings of the system.
We are currently trying 
to further characterize the nature (order) of the transition of the dynamical model.
We have furthermore shown that the onset of friction and the pinning of the system are related 
but distinct phenomena occurring in general at different times, which we have been able to identify. 

\section*{Acknowledgements}

We would like to thank Ted Janssen for interesting discussions and suggestions.

\end{multicols}

\appendix
\section*{}
\label{gam_d}

We are going to provide in this appendix an explicit derivation that $\dot{p}_{\phi}=0$.
Let us consider the first term on the right hand side of Eq. (\ref{gam}):
\begin{equation}
L = -i\sum_n nqx_{-nq}\dot{x}_{nq}
\label{gal}
\end{equation}
In order to simplify the notation, we will 
adopt from now on the following convention: $ nq \equiv \kappa$.
Take the derivative of Eq. (\ref{gal}):
\begin{equation}
\dot{L}=-i\sum_{\kappa} \kappa x_{-\kappa}\ddot{x}_{\kappa} -i\sum_{\kappa} \kappa\dot{x}_{-\kappa}\dot{x}_{\kappa}
\label{derang}
\end{equation}
We can immediately see that for symmetry reasons the second term cancels.
Let us now take the equation for $\ddot{x}_k$ as follows from the Euler-Lagrange 
equations:
\[
\ddot{x}_{\kappa}=-\omega^2_{\kappa} x_{\kappa} + \frac{\lambda}{2}\sum_{m=0}^\infty 
\frac{(i2\pi)^m}{m!}\sum_{\kappa_1\ldots \kappa_m}\left[e^{i2\pi Q}x_{\kappa_1}\cdots 
x_{\kappa_m}\delta_{\kappa_1+\cdots
+\kappa_m,-q+\kappa}\right. 
\]
\begin{equation}
\left. + (-1)^m
e^{-i2\pi Q} x_{\kappa_1} \cdots x_{\kappa_m}\delta_{\kappa_1+\cdots +\kappa_m,q+\kappa} \right]
\label{expans1}
\end{equation}
and the equation for the CM motion:
\[
\ddot{Q} = \frac{\lambda}{2}\sum_{m=1}^\infty\frac{(i2\pi)^m}{m!}\sum_{\kappa_1\ldots \kappa_m}\left[e^{i2\pi Q}x_{\kappa_1}\cdots 
x_{\kappa_m}\delta_{\kappa_1+\cdots
+\kappa_m,-q}\right. 
\]
\begin{equation}
\left. + (-1)^m
e^{-i2\pi Q} x_{\kappa_1} \cdots x_{\kappa_m}\delta_{\kappa_1+\cdots +\kappa_m,q} \right]
\label{expansQ}
\end{equation}
Let us insert Eq. (\ref{expans1}) in Eq. (\ref{derang}). We get:
\begin{equation}
\dot{L}=-i\left\{\frac{\lambda}{2}\sum_{m=0}^{\infty}\frac{(i2\pi)^m}{m!}
  \sum_{\kappa,\kappa_1,\ldots,\kappa_m}\left[e^{i2\pi Q}\kappa x_{-\kappa}x_{\kappa_1}\cdots x_{\kappa_m}
  \delta_{\kappa_1 + \cdots + \kappa_m, -q+\kappa} + ... \right]\right\}
\label{derang1}
\end{equation}
(Note that the first term in Eq. (\ref{expans1}) cancels for the same symmetry 
reasons given above). We have for simplicity explicitly written down only the first part
of the expression in square brackets, since
we treat the second part exactly in the same way.
Rearranging the delta function and applying
the following symmetry transformation:
\begin{equation}
\kappa x_{-\kappa}\delta_{-\kappa} \rightarrow -\kappa x_{\kappa}\delta_{\kappa}
\label{transf2}
\end{equation}
Eq. (\ref{derang1}) becomes:
\begin{equation}
\dot{L}=-i\left\{\frac{\lambda}{2}\sum_{m=0}^{\infty}\frac{(i2\pi)^m}{m!}
  \sum_{\kappa,\kappa_1,\ldots,\kappa_m}\left[e^{i2\pi Q}(-\kappa)x_{\kappa}x_{\kappa_1}\cdots x_{\kappa_m}
  \delta_{\kappa+ \kappa_1 + \cdots + \kappa_m, -q} + ... \right]\right\}
\label{derang2}
\end{equation}
Because there is no preferential order in 
the $\kappa$-summation, the following equality holds:
\begin{equation}
(-\kappa)x_{\kappa}x_{\kappa_1}\cdots x_{\kappa_m} = (-\kappa_1)x_{\kappa_1}x_{\kappa}\cdots x_{\kappa_m} = 
\cdots = (-\kappa_m)x_{\kappa_m}x_{\kappa}\cdots x_{\kappa_{m-1}}
\label{perm}
\end{equation}
There are $(m+1)$ possibilities, thus
we can make the substitution:
\begin{equation}
\kappa = \frac{\kappa + \kappa_1 + \cdots + \kappa_m}{m+1}
\label{subst}
\end{equation}
Therefore, Eq. (\ref{derang1}) becomes:
\begin{equation}
\dot{L}=-i\left\{\frac{\lambda}{2}\sum_{m=0}^{\infty}\frac{(i2\pi)^m}{m!}
  \sum_{\kappa,\kappa_1,\ldots,\kappa_m}\left[\left(-\frac{\kappa + \kappa_1 + \cdots + \kappa_m}{m+1}\right)
   e^{i2\pi Q}x_{\kappa}x_{\kappa_1}\cdots x_{\kappa_m}
  \delta_{\kappa + \kappa_1 + \cdots + \kappa_m, -q} + \cdots \right]\right\}
\label{derang3}
\end{equation}
Now, under the assumption that there is no Umklapp $\kappa_1 + \cdots \kappa_m = q$ and
can be taken outside the summation. Hence, we get: 
\begin{equation}
\dot{L}=-\frac{q}{2\pi}\left\{\frac{\lambda}{2}\sum_{m=0}^{\infty}\frac{(i2\pi)^{m+1}}{(m+1)!}
  \sum_{\kappa, \kappa_1, \cdots, \kappa_m}e^{i2\pi Q}x_{\kappa_1}\cdots x_{\kappa_m}
  \delta_{\kappa + \kappa_1 + \cdots + \kappa_m, -q} + \cdots \right\}
\label{derang4}
\end{equation}
The expression in parenthesis is precisely Eq. (\ref{expansQ}) for $\ddot{Q}$ multiplied by $q/2\pi$. Hence, we find:
\begin{equation}
\dot{p}_{\phi}=\dot{L} + \frac{q}{2\pi}\ddot{Q} = 0
\label{con}
\end{equation}

\end{document}